\documentclass[aps,pra,twocolumn,superscriptaddress,floatfix,amsmath,amssymb]{revtex4-1}
\usepackage[utf8]{inputenc}
\usepackage{mathtools}
\usepackage[english]{babel}
\usepackage{hyperref}
\usepackage{float}
\usepackage{graphics}
\usepackage{xcolor}
\usepackage{bbold}
\usepackage[dvips]{graphicx} 
\usepackage[graphicx]{realboxes}

\newcommand* {\bra}[1]{\ensuremath{\langle {#1} |}}
\newcommand* {\ket}[1]{\ensuremath{| {#1} \rangle}}
\newcommand{\ketbra}[2]{|{#1}\rangle\langle{#2}|}

\newcommand{\hbrho}{\boldsymbol{\rho}}

\begin{document}
	
	\title{
		Performance of 
  entanglement purification including 
		maximally entangled mixed states
	}
		\author{Juan Mauricio Torres}
	\email{jmtorres@ifuap.buap.mx}
	\affiliation{Instituto de F\'isica, Benem\'erita Universidad Aut\'onoma de Puebla, C.P. 72570, Puebla, Mexico.}
 	\author{J\'ozsef Zsolt Bern\'ad}
	\affiliation{Peter Gr\"unberg Institute (PGI-8), Forschungszentrum J\"ulich, D-52425 J\"ulich, Germany}
	\author{Roc\'io G\'omez-Rosas}
		\affiliation{Instituto de F\'isica, Benem\'erita Universidad Aut\'onoma de Puebla, C.P. 72570, Puebla, Mexico.}
	\date{\today}
	
\begin{abstract}
Entanglement between distant quantum systems is a critical resource for implementing quantum communication. This property is affected by external agents and can be restored by employing efficient entanglement purification protocols. In this work, we propose an entanglement purification protocol based on two entangling two-qubit operations that replace the usual controlled-NOT (CNOT) gate. These operations arise from a generalized quantum measurement and can be understood as measurement operators in a positive operator-valued measure (POVM). Furthermore, two variants of the core protocol are introduced and shown to be more practical in certain scenarios. The performance of the protocols is studied in terms of the overall success probability of reaching a Bell state and the number of purifiable states. Based on rank-two states, we can obtain analytical expressions for the success probability that we extend and refine using numerical calculations to the case of maximally entangled states (MEMS). We also consider more general rank-three states to show that our procedure is in general more convenient compared to purification protocols based on Bell diagonal states. Finally, we test the protocols using initial random states. In all cases, we find a larger performance and larger amount of purifiable states using our schemes compared to the  CNOT-based purification protocol.

\end{abstract}
\maketitle
\section{Introduction}
\label{Int}

Quantum technologies have experienced in the last two decades promising advances and the developed quantum protocols use entanglement as a key resource, which is distributed among distant or nearby nodes of a network of quantum systems. Applications vary from quantum communication \cite{Sangouard, Froehlich, Pirandola2020}, simulations \cite{Monroe}, computation \cite{Albash} to atomic clocks \cite{Ludlow} and material qubits hold great promise for these research fields. Experimentally, there is also ongoing progress, and for example entanglement between remote neutral atoms \cite{Ritter}, NV centers \cite{Hanson}, trapped ions \cite{Hucul}, and superconducting qubits \cite{Leung} was achieved. However, the created entangled pairs are not maximally entangled, e.g. $0.793 \pm 0.003$ fidelity with respect to a Bell state \cite{Leung}, because, in physical reality, quantum networks are subject to information loss and thus one of the main tasks is to protect quantum information and its processing. Furthermore, quantum information cannot be cloned or amplified without changing its quantum nature \cite{Dieks, Zurek}. To protect quantum information one may use quantum error correction \cite{Devitt} or entanglement purification. The latter is the subject of this paper.

Quantum teleportation assisted by entanglement purification  plays a fundamental role in the implementations of quantum technologies, in particular in quantum network-based communication. The first recurrence protocol was introduced by Bennett and collaborators \cite{Bennett1996}, which relies on Werner states \cite{Werner1989}. A more general protocol was introduced by Deutsch {\it et al.}  (DEJMPS) \cite{Deutsch1996} and it is based on Bell diagonal states, a family of 
three-parameter states. These approaches use local random transformations to convert a general two-qubit state to the one required by the protocol, which wastes useful entanglement \cite{Bennett1996b, Horodecki}. The other central element is the use of the controlled-NOT (CNOT) gate, similar to the XOR (exclusive OR) gate in  classical computer science. While the XOR gate is effectively implemented by semiconductor-based classical circuits  \cite{Rabaey}, the CNOT gate is subject to imperfect gate fidelities and further trade-offs in performances \cite{Blatt, Brien, Zajac, Mills}. However, there is another approach, when one does not focus on abstract entangling quantum gates but instead investigates the emergence of physical interaction-based entangling operations, which appear naturally, although they are not always unitary \cite{Gonta, Bernad2016, Rosas2021}. This approach leads to alternative entanglement purification protocols, which yield the same results for Werner states \cite{Bernad2016}, but for general two-qubit states their performance may vary and investigations of optimal strategies are also required \cite{Krastanov, Preti}.   

In this paper, we introduce improved  alternatives of these recurrence protocols and compare their performance with the DEJMPS protocol. We present the mathematical description of the iterations and show that it is beneficial in the first round as it maps any input state onto an $X$-state \cite{Rau} that is maintained in further iterations. We briefly analyze the convergence properties of the protocols and apply the results to Bell diagonal and maximally entangled mixed states (MEMS) \cite{Ishizaka2000}, which have a high degree of entanglement for a given purity. MEMS of one type hold a particular significance for performance evaluations because they can be purified in one step with both unitary map-based \cite{Bennett1996b} and alternative protocols \cite{Torres2016}. Furthermore, we propose and investigate a generalized version of MEMS to show that finite iteration steps of the purification protocol are enough to obtain maximally entangled states without using any additional available entanglement \cite{Eisert, Duer2021}, which is in contrast to the countable infinite number of iteration steps required by the original protocols. Finally, performances on the average required qubit pairs and the percentage of the purifiable states among all possible initial states are numerically investigated. Thus, the main aim of 
the paper is to show that physical interactions-based entangling operations can yield better performances than the original CNOT-based entanglement purification protocols that are also done with idealized quantum hardware. 

This paper is organized as follows. We introduce the core entanglement purification protocol in Sec. \ref{sec:MainProtocol}. In Sec. \ref{sec:BellDiagonal}, we show its relation with DEJMPS protocol and evaluate analytically the success probability for rank-two states in the Bell basis. In Sec. 
\ref{sec:ProtocolHadamard}, we present a modification of the protocol assisted by Hadamard gates presenting two purification paths, were in one of them a Bell diagonal state can be obtained during the first purification step, while in the other path the states remain in the $X$-state form.
In Sec. \ref{sec:SecondProtocolHadamard},
we present a second modification where we manipulate the outcomes in order to continue the process in a Bell diagonal form. We concentrate on the purification of maximally entangled mixed states in Sec. \ref{sec:MEMS}. Based on the form of MEMS, we extend our study to other rank-three states in Sec. \ref{sec:Rank3States}.
 Numerical results for generic random states are presented in Sec. \ref{sec:Numerical}, and we draw our conclusions  in Sec. \ref{sec:Conclusions}. Finally,  a detailed discussion of the discrete probability distribution of the number of successes in an iterated protocol is presented in the Appendix.

 \section{Entanglement purification (M2) protocol}
 \label{sec:MainProtocol}
 In this work we consider  an improved version of the entanglement
 purification protocol introduced in \cite{Bernad2016} which was based on a non-unitary quantum operation,
 realizable in cavity QED systems, in place of the usual
 CNOT gate.  Here we extend the purification protocol using the following pair of two-qubit operations
 \begin{equation}
 M_\pm=\ket{\Psi^{\pm}}\bra{\Psi^{\pm}}+
 	\ket{\Phi^{\pm}}\bra{\Phi^{\pm}},
 \end{equation}
that we have expressed in terms of the Bell states
\begin{align}
\ket{\Psi^\pm}=\frac{\ket{01}\pm\ket{10}}{\sqrt2} ,\quad
	\ket{\Phi^\pm}=\frac{\ket{00}\pm \ket{11}}{\sqrt2}.
	\label{Bellstates}
\end{align}
As the present improved protocol is based on two quantum operations $M_\pm$ we will label it, for convenience, as M2 protocol.
Previously in \cite{Bernad2016}, we employed only $M_-$ motivated by the fact that also the seminal entanglement purification
protocols are intrinsically
probabilistic, as their success depends on a measurement outcome
in the computational basis, which takes place after the application of the CNOT gate. In contrast, $M_-$  can be considered as a measurement operator that is in addition an entangling non-unitary operation. To complete the measurement scheme, one 
requires the additional operator $M_+$ that has not been studied before
in the context of entanglement purification. Recently, it has been shown
that implementations of both $M_+$ and $M_-$ are possible in physical systems \cite{Rosas2021,Gonzalez2019}
and can be understood as elements of a POVM \cite{Gonzalez2019}. 
As they are rank-two projectors in a two-qubit system, it follows that $M_\pm^\dagger M_\pm=M_\pm^2=M_\pm$ and $M_+^2+M_-^2=M_++M_-=\mathbb{1}$, i.e., they sum to the identity matrix. With this in consideration, it is plausible to infer that the
overall success probability can be enhanced by incorporating $M_+$ into the protocol. We shall demonstrate that this is indeed the case.

Let us briefly recapitulate all the required steps in each iteration of the
considered recurrence entanglement purification protocol, and explain how the operations $M_\pm$ enter into play. Two exact copies
of entangled mixed states shared by distant nodes $A$ and $B$ are initially found to be in the state
	\begin{equation}
	\hbrho=\rho^{A_1,B_1}\otimes \rho^{A_2,B_2}.
	\label{rhoAB4}
\end{equation}
The main idea of the purification scheme is to trade these two entangled mixed states with a single state having a higher degree of entanglement. As the qubit-pairs are shared by spatially separated nodes, only local operations in those locations and classical communication between them are possible. Usually, a large number of copies of these two-qubit states is required, but not always \cite{Bennett1996b, Torres2016}, to repeat the process until a state as close as possible to a maximally entangled state, in our case a Bell state, is reached. 

\subsection{Purification process in one iteration}
\label{sec:oneit}

Each iteration of the entanglement purification protocol can be summarized in three parts as follows. \\
{\bf (I)} The two-qubit operations $M_\pm$ 
are  bilaterally applied at each node transforming the initial four-qubit $\hbrho$ state into
	\begin{equation}
	\hbrho^\pm=
	\Pi_\pm \hbrho \Pi^\dagger_\pm 
	,\quad 
	\Pi_\pm=M_\pm^{A_1,A_2} \otimes M_\pm^{B_1,B_2},
	\label{maprho}
\end{equation}
where we have introduced the operation $\Pi_\pm$, which  is non-unitary,  
and therefore
$\hbrho^\pm$ is in general not normalized. \\
{\bf (II)} Qubits $A_2$ and $B_2$ are measured in the computational basis and discarded in order to leave qubits 
$A_1$ and $B_1$ in the state
 \begin{equation}
 	\tilde\rho^{\pm A_1,B_1}=q_\pm^{-1}\,{\rm Tr}_{A_2,B_2}\left\{
 	\hbrho^\pm
 	(I^{A_1,B_1} \otimes \ketbra{jk}{jk}^{A_2,B_2})
 	\right\},
 \end{equation}
where $\ket{jk}^{A_2,B_2}\equiv\ket{j}^{A_2}\ket{k}^{B_2}$ with $j,k\in\{0,1\}$ and $I^{A_1,B_1}$ is the identity map on qubits $A_1$ and $B_1$. The normalization of the state is given by 
\begin{equation}
	q_\pm={\rm Tr}_{A_2,B_2}\left\{
	\Pi_\pm \hbrho\Pi_\pm^\dagger
	(I^{A_1,B_1} \otimes \ketbra{jk}{jk}^{A_2,B_2})
	\right\}.
 \label{eq:proq}
\end{equation}
This normalization factor is the success probability of the non-unitary process
involving the implementation of $\Pi_\pm$ and the measurement of qubits $A_2$ and $B_2$ in the
computational basis. \\
{\bf (III)} Finally, depending on the outcome of the measurement,
single qubit gates are applied to qubits $A_1$ and $B_1$ in the following way
	\begin{align}
		\rho^{\pm A_1,B_1}=
		\left(V_j^{A_1} \otimes V_{k+1}^{B_1}\right)
\tilde\rho^{\pm A_1,B_1}
		\left(V_j^{A_1} \otimes V_{k+1}^{B_1}\right)^\dagger
	\end{align}
with the single-qubit gate 
$V_j=\ket{1}\bra{j\oplus 1}+i\ket{0}\bra{j}$ expressed in terms of $\oplus$, which denotes the sum modulo $2$. This last step does not affect entanglement, however, it is necessary to obtain 
always the same form of the final state. 
	
After the three-step process and depending on the use of $M_-$ or $M_+$, one can obtain a closed form expression for the matrix elements of the output
density matrix $\rho^\pm$ describing only qubits $A_1$ and $B_1$. We will drop these qubit labels at this point as we are left only with one qubit pair after the other one is measured. In the standard Bell basis ordered in the form 
\begin{equation}
    \{ 1 \leftrightarrow \ket{\Psi^-}, 2 \leftrightarrow \ket{\Phi^-}, 3 \leftrightarrow \ket{\Phi^+}, 4 \leftrightarrow \ket{\Psi^+}\},
\end{equation}
the matrix elements of the output state are given by 
		\begin{align}
		\rho^\pm_{11}&=\frac{\rho^2_{11}+\rho^2_{22}\pm\rho^2_{12}\pm\rho^2_{21}}{2q_\pm}, \quad 
		\rho^\pm_{22}=\frac{\rho_{33}\rho_{44}\pm|\rho_{34}|^2}{q\pm}, \nonumber \\
		\rho^\pm_{44}&=\frac{\rho^2_{33}+\rho^2_{44}\pm\rho^2_{34}\pm\rho^2_{43}}{2q_\pm},
		\quad
		\rho^\pm_{33}=\frac{\rho_{11}\rho_{22}\pm|\rho_{12}|^2}{q_\pm}, 
		\nonumber  \\
		\rho^\pm_{14}&=\frac{\rho^2_{14}+\rho^2_{23}\pm\rho^2_{13}\pm\rho^2_{24}}{2q_\pm},  
		\quad \rho^\pm_{23}=\frac{\rho^*_{23}\rho^*_{14}\pm\rho^*_{13}r^*_{24}}{q_\pm}.
		\label{map1}
	\end{align}
Together with their transposed elements that can be obtained by complex conjugation, e.g. 
$\rho_{41}^\pm=(\rho_{14}^{\pm})^\ast$, these are the only nonzero elements whenever $M_-$ is bilaterally applied. In this case, the two-qubit system
 attains an  $X$-state form, where $\rho^-_{12}=\rho^-_{13}=\rho^-_{41}=\rho^-_{42}=0$.  This family of two-qubit states was already encountered in \cite{Bernad2016}. When $M_+$ is implemented
 at both nodes, one also encounters the following potentially nonzero elements
	\begin{align}
	\rho_{12}^+&=\frac{\rho_{13}\rho_{14}+\rho_{23}\rho_{24}}{iq_\pm},  \quad
		\rho_{13}^+&=\frac{\rho_{11}\rho_{12}+\rho_{22}\rho_{21}}{iq_\pm},  \nonumber\\
			\rho_{42}^+&=\frac{\rho_{44}\rho_{43}+\rho_{33}\rho_{34}}{iq_\pm},  \quad
		\rho_{43}^+&=\frac{\rho_{31}\rho_{32}+\rho_{41}\rho_{42}}{iq_\pm}.
	\label{map2plus}
\end{align}
Note, however, that if  $\rho$ is already in an $X$-state, then all these contributions in Eq.~\eqref{map2plus} vanish and the elements in Eq.~\eqref{map1} are equal in both cases, i.e., whenever $M_-$ or $M_+$ are bilaterally applied. 
The success probability  can be evaluated for the general case from Eq. \eqref{eq:proq}  and is given by
	\begin{align}
	q_\pm(\rho)=&\frac{
		(\rho_{11}+\rho_{22})^2+(\rho_{33}+\rho_{44}) ^2}{2}\nonumber\\
	&\qquad\qquad \pm2{\rm Re}[\rho_{12}]^2\pm2{\rm Re}[\rho_{34}]^2.
\end{align}
We have that $q_-+q_+<1$, as we are still missing two possible outcomes of the process that
are given when $M_\mp$ is applied at node $A$ and correspondingly $M_\pm$ at node $B$.  If this is the case, one
obtains a general density matrix with very different elements. The success probability for this asymmetric 
process is then given by 
$r_\pm=(\rho_{11}+\rho_{22})(\rho_{33}+\rho_{44})\pm(\rho_{12}+\rho_{21})(\rho_{34}+\rho_{43})$.  Taking into account
all probable processes, one finds that $q_-+q_++r_-+r_+=1$, i.e., the probabilities sum up to one as expected. In order to keep the form
of the  output state in every iteration round of the protocol, we only consider processes where the operations on both sides coincide. 

By iterating the map in Eq. \eqref{maprho} for the minus sign, one is able 
to asymptotically generate a Bell state depending on the initial conditions.
As shown in \cite{Torres2016}, if the condition
\begin{equation}
\label{eq:conditionpsim}
  (2\rho_{11}-1)(1-2\rho_{22})>-(2{\rm Im}[\rho_{12}])^2-(2{\rm Re}[\rho_{34}])^2  
\end{equation}
is fulfilled, the Bell state $\ket{\Psi^-}$ is generated asymptotically.  
Whenever we have 
\begin{equation}
\label{eq:conditionpsip}
(2\rho_{33}-1)(1-2\rho_{44})>-(2{\rm Im}[\rho_{34}])^2-(2{\rm Re}[\rho_{12}])^2, 
\end{equation}
then state $\ket{\Psi^+}$
is approached. If none of these conditions is met, the maximally mixed state is reached.

\subsection{Further iterations and overall success probability}

In the case of several iterations, it is assumed again that before the next round of purification, the successfully obtained identical states are paired up again. This should be kept in mind for the rest of the paper because the maps of the iterations do not explicitly reveal this assumption. After a first successful iteration with the bilateral $M_-$ two-qubit operation, the states are left in an $X$-state  form
\begin{equation}
\label{eq:XStates}
\rho'=
	\begin{pmatrix}
		\rho'_{11}& 0&0&\rho'_{14}\\
		0&\rho'_{22}&\rho'_{23}&0\\
		0&\rho'_{32}&\rho'_{33}&0\\
		\rho'_{11}& 0&0&\rho'_{14}
	\end{pmatrix}
\end{equation}
with $\rho'=\rho^-$. 
Therefore, it follows from Eqs. \eqref{map1} and \eqref{map2plus}
that one can use both operations $M_-$ and $M_+$ in further iterations, as both  preserve the $X$-state form of the output state possessing the following nonzero matrix elements
\begin{align}
	\rho''_{11}&=\frac{\rho'^2_{11}+\rho'^2_{22}}{p(\rho')}, 
	\quad 
		\rho''_{44}=\frac{\rho'^2_{33}+\rho'^2_{44}}{p(\rho')},
	\quad
		\rho_{14}''=\frac{\rho'^2_{14}+\rho'^2_{23}}{p(\rho')},  
	\nonumber \\
	\rho''_{22}&=2\frac{\rho'_{33}\rho'_{44}}{p(\rho')}, 
	\quad
	\rho''_{33}=2\frac{\rho'_{11}\rho'_{22}}{p(\rho')},  
	\quad
	\rho''_{23}=2\frac{\rho'^*_{23}\rho'^*_{14}}{p(\rho')},
	\label{map2}
\end{align}
where the success probability $p(\rho')$ is given as
a function of the matrix elements:
	\begin{equation}
p(\rho)=
	(\rho_{11}+\rho_{22})^2+(\rho_{33}+\rho_{44}) ^2.
	\label{eq:succprob2}
\end{equation}
Thus, the iterations of the purification process can be sketched in the following way 
\begin{equation}
\rho \xrightarrow{M_-} \rho^-=\rho'\xrightarrow{M_\pm}\rho''\xrightarrow{M_\pm}\rho'''\dots
\rho^{(l)}\xrightarrow{M_\pm}\rho^{(l+1)}, \nonumber
\end{equation}
where the initial state $\rho$ is an arbitrary two-qubit state. The arrow 
$\xrightarrow{M_\pm}$ represents one iteration process composed of the steps {\bf (I)} to {\bf (III)} as explained in section \ref{sec:oneit}. The outputs of
further applications of the map are expressed using primes, e.g., $\rho''$ represents
the state after the second iteration that we also denote as $\rho^{(2)}$. Note
that in the first iteration, only the operation $M_-$ is considered. In a sequence of many iterations, the whole process can be approximated by a single Bernoulli trial with a success probability obtained by multiplying the success probabilities of all iterations, see Appendix \ref{AppendixA}. This overall success probability can describe the process of purifying highly entangled states from a large number of input states with a high number of iterations of the protocols, a case, which will be studied throughout the whole paper. As we have shown in Appendix \ref{AppendixA}, any other particular situation requires a more detailed calculation. Thus, based on Appendix \ref{AppendixA}, the success probability of the whole process reads 
\begin{equation}
\label{eq:probM2}
	P=q_-(\rho)p(\rho')p(\rho'')\dots=q_- (\rho)\prod_{l=1}^\infty
	p(\rho^{(l)}).
\end{equation}
Note, however, that  for initial 
states $\rho$  in an $X$-state form, one can also
employ $M_+$ in the first round, and
the overall success probability takes the form
\begin{equation}
\label{eq:probM2X}
	P_X=p(\rho)p(\rho')p(\rho'')\dots =\prod_{l=0}^\infty
	p(\rho^{(l)}).
\end{equation}
This is the particular case, for instance, of the Bell diagonal states that we consider in the next section.

\section{Bell diagonal states}
\label{sec:BellDiagonal}

\subsection{Connection with the DEJMPS protocol}

The recurrence purification protocol introduced by Deutsch and collaborators (DEJMPS) in \cite{Deutsch1996} is only valid for Bell diagonal states. It has the  same map as the diagonal entries in Eq. \eqref{map2}, but with the Bell basis ordered differently. To get the same map, one can consider
the following single-qubit unitary rotations
in terms of the Hadamard gate $H$
\begin{align*}
	\{\ket{\Psi^-},\ket{\Phi^-},\ket{\Phi^+},\ket{\Psi^+}\}
	\xrightarrow{H^{\otimes2}}
	\{-\ket{\Psi^-},\ket{\Psi^+},\ket{\Phi^+},\ket{\Phi^-}\}
	\\
	\xrightarrow{\mathbb{1}\otimes\sigma_x\sigma_z}
	\{\ket{\Phi^+},-\ket{\Phi^-},\ket{\Psi^-},\ket{\Psi^+}\},
\end{align*}
where $\sigma_x$ and $\sigma_y$ are the Pauli matrices \cite{Nielsen2000}. These rotations map Bell states
onto one another. If one uses the gate
$H\otimes(\sigma_x\sigma_yH)$ before and after our protocol, then for Bell diagonal states our map corresponds to the one in the DEJMPS protocol \cite{Deutsch1996}. The success probability is the same as in Eq. \eqref{eq:succprob2}. Therefore, our scheme, based on the operations $M_+$ and $M_-$, is equivalent to the 
DEJMPS protocol for Bell diagonal states. 

In cases of general states,
twirling operations that diminish the entanglement are necessary to bring 
arbitrary states to the Bell diagonal form, see the DEJMPS protocol \cite{Deutsch1996}. In contrast, this step is not required
by our protocol. This is reflected in the fact that in our protocol, conditions for a state to be purified are relaxed as noted in Eqs. \eqref{eq:conditionpsim} and \eqref{eq:conditionpsip}, which also include the requirements of the DEJMPS protocol, where one of the fidelities of the input state with respect to a Bell state must be larger than $0.5$ \cite{Macchiavello}. For $X$-states, the map and the success probability are the same as their counterparts in the DEJMPS protocol, because the diagonal entries are decoupled from the off-diagonal ones as noted in Eq. \eqref{map2}. However, in our case, no preparatory step is needed to bring the state to the Bell diagonal form.

\subsection{Overall success probability for rank-two states}
\label{rank-two}

Rank-two states in the Bell basis, i.e., the convex combination of two Bell state projectors, 
allow us to calculate analytically the overall success probability.  There are six possible ways to combine
two different Bell states to form a Bell diagonal state with only two nonzero eigenvalues. 
The convex combination of  $\ketbra{\Psi^-}{\Psi^-}$
and $\ketbra{\Psi^+}{\Psi^+}$ is important to analyze, as it retains its form after successive iterations of the protocol, as follows
from Eq. \eqref{map2}. Also, based on Eq. \eqref{map2}, one can observe that the other five combinations reach this form 
after a certain number of iterations. This happens for one iteration for the combination involving the states
$\ketbra{\Phi^\pm}{\Phi^\pm}$.
For the moment, we will restrict the analysis to the following family of states that retains its form after successive iterations of the protocol, namely
\begin{equation}
	\rho^{(n)}=a_n\ketbra{\Psi^\mp}{\Psi^\mp}+(1-a_n)\ketbra{\Psi^\pm}{\Psi^\pm}.
	\label{eq:rank2}
\end{equation}
For symmetry reasons and without loss of generality, one can take $1/2<a_0\le 1$, where $a_0$ will be the initial condition in the map representing the protocol. The exact expression for  $a_n$ after $n$ iterations
of the map can be evaluated in an analytical form. From Eq. \eqref{map2}, we find the first-order recurrence
relation between values in successive iterations
\begin{equation}
	a_{n}=\frac{a_{n-1}^2}{a_{n-1}^2+(1-a_{n-1})^2}=\frac{a_0^{2^n}}{a_0^{2^n}+(1-a_0)^{2^n}},
	\label{eq:rank2map}
\end{equation}
where the solution of the recurrence relation is obtained
by introducing the auxiliary 
variable 
\begin{equation}
	b_n=a_n^{-1}-1=b_{n-1}^2=b_0^{2^n}.
\end{equation}
The success probability for the $n$-th iteration is evaluated from the coefficients of the previous iteration as
$p_{n}=a_{n-1}^2+(1-a_{n-1})^2=(1+b_{n})/(1+b_{n-1})^2$. The overall success probability after $n$ iterations is given by 
\begin{equation}
	P_n=\prod_{k=1}^n p_k=\prod_{k=1}^n 
	\frac{1+b_0^{2^{k}}}{(1+b_0^{2^{k-1}})^2}=
	\frac{1+b_0^{2^{n}}}{1-b_0^{2^{n}}}\frac{1-b_0}{1+b_0}.
\end{equation}
The product has been evaluated using the following relation with a geometric sum
\begin{equation}
\prod_{k=0}^{n-1}(1+x^{2^k})=
\sum_{k=0}^{2^{n}-1}x^k=
\frac{1-x^{2^{n}}}{1-x}.
\end{equation}
Therefore, the overall success probability to purify a Bell state leads to the following result
\begin{equation}
	P=\lim_{n\to\infty} P_n=\frac{1-b_0}{1+b_0}=2a_0-1=C,
	\label{eq:rank2Pinf}
\end{equation}
where $C=2a_0-1$ is the concurrence  of the initial state, an entanglement measure introduced in \cite{Wootters1998} that takes unite value for maximally entangled states and vanishes for separable states. 
A convex combination of $\ket{\Psi^\mp}$ and $\ket{\Phi^\pm}$ leads after one iteration to a state of the form in Eq. \eqref{eq:rank2} and eventually the state $\ket{\Psi^\mp}$ is purified. More importantly, the map in Eq. \eqref{eq:rank2map}
holds for this case as well, therefore the success probability is also given by $C$ as in Eq. \eqref{eq:rank2Pinf}.

The situation is different for the convex combination of 
$\ket{\Psi^\mp}$ and $\ket{\Phi^\mp}$. In this case, two iterations are necessary to reach a state of the form of Eq. \eqref{eq:rank2}. The first iteration leads 
$a_1=a_0^2+(1-a_0)^2$, however, this happens with success probability $p_1=1$. The map in Eq. \eqref{eq:rank2} holds for $n\ge2$, therefore the overall success probability is given by $P=2a_1-1=C_1$ or $P=(2a_0-1)^2=C^2$, i.e., the initial concurrence squared. 

It is worth mentioning that these type of states can be purified for any non-vanishing value of the initial concurrence with a finite probability. This is true for both cases  DEJMPS protocol and our M2 protocol. We will show that this analytical result will be quite helpful for the calculation of the success probability of other type of  states.  

\section{Protocol with Hadamard gates (M2H)}
\label{sec:ProtocolHadamard}

In this section, we introduce a modification of the M2 protocol using Hadamard gates  $H$  acting on each qubit. We name this variant as M2H protocol.  Starting from an $X$-state $\rho'$, the states are transformed to a block diagonal form using $H^{\otimes 2}$. We will show that, after this procedure, an iteration of the protocol with $M_-$ brings these states into a Bell diagonal form.

Arbitrary states can be transformed into a Bell diagonal
state using random unitary or twirling operations in the form $\rho\to \sum_j U_j\rho U_j^\dagger$ \cite{Bennett1996b}. This process is detrimental to the initial entanglement. 
An important feature of the present protocol is its possibility 
to bring general two-qubit states to a 
Bell diagonal form without the
typical twirling operations required in the 
DEJMPS protocol. Furthermore, in this process,  the states increase their degree of entanglement.  For this, it is required that the state is brought to an $X$-state form. Afterwards,
Hadamard gates are applied to each qubit  
bringing the state to a Bell diagonal form
in such a way that 
\begin{equation}
\label{eq:rhoHadmard}
H^{\otimes 2}\rho'
H^{\otimes 2}=
	\begin{pmatrix}
		\rho'_{11}& -\rho'_{14}&0&0\\
		-\rho'_{41}&\rho'_{44}&0&0\\
		0&0&\rho'_{33}&\rho'_{32}\\
		0&0&\rho'_{23}&\rho'_{22}
	\end{pmatrix}.
\end{equation}
An iteration of the protocol using $M_\pm$ with this state yields the diagonal entries
\begin{align}
\label{eq:rhodiagHad}
	\rho^{\pm,1}_{11}&=
	(\rho_{11}'^2+\rho_{44}'^2\pm\rho_{14}'^2\pm\rho_{41}'^2)/2Q_\pm(\rho'),\nonumber\\
	\rho^{\pm,1}_{22}&=	(\rho_{22}'\rho_{33}'\pm|\rho_{23}'|^2)/Q_\pm(\rho'),\nonumber\\
	\rho^{\pm,1}_{33}&=	(\rho_{11}'\rho_{44}'\pm|\rho_{14}'|^2)/Q_\pm(\rho'),\nonumber\\
	\rho^{\pm,1}_{44}&=	(\rho_{22}'^2+\rho_{33}'^2\pm\rho_{23}'^2\pm\rho_{32}'^2)/2Q_\pm(\rho'),
\end{align}
with success probability
\begin{align}
	\label{eq:probHadMs}Q_\pm(\rho)=&q_\pm(H^{\otimes2}\rho H^{\otimes 2})=
	\frac{
		(\rho_{11}+\rho_{44})^2+(\rho_{22}+\rho_{33})^2}
	{2}
	\nonumber\\
 &\qquad\qquad \pm2{\rm Re}[\rho_{14}]^2\pm2{\rm Re}[\rho_{23}]^2.
\end{align} 
Then, the state is transformed into Bell diagonal form under the M2 protocol with $M_-$ as
\begin{equation}
\label{eq:XHadMminus}
H^{\otimes 2}\rho'H^{\otimes 2}
\xrightarrow{M_-}\rho^{-,1}=
	\begin{pmatrix}
		\rho^{-,1}_{11}&0&0&0\\
		0&\rho^{-,1}_{22}&0&0\\
	0&0&\rho^{-,1}_{33}&0\\
		0&0&0&\rho^{-,1}_{44}
	\end{pmatrix}.
\end{equation}
This Bell diagonal form will be preserved during subsequent iterations using either $M_-$ or $M_+$.
This is not the case, when $M_+$ is initially applied, where the state is mapped according to
\begin{equation}
\label{eq:XHadMplus}
H^{\otimes 2}\rho'H^{\otimes 2}
\xrightarrow{M_+}\rho^{+,1}=
	\begin{pmatrix}		\rho^{+,1}_{11}&0&\rho^{+,1}_{13}&0\\	0&\rho^{+,1}_{22}&0&\rho^{+,1}_{24}\\
	\rho^{+,1}_{31}&0&\rho^{+,1}_{33}&0\\
		0&\rho^{+,1}_{42}&0&\rho^{+,1}_{44}
	\end{pmatrix}.
\end{equation}
The off-diagonal elements can be evaluated from Eq. \eqref{map2plus} with the initial state $H^{\otimes 2}\rho'H^{\otimes 2}$ in \eqref{eq:rhoHadmard} and they read
\begin{equation}
    \label{eq:rhooffdiagHad}
    \rho_{13}^{+,1}=
i\frac{\rho_{11}'\rho_{14}'+\rho_{44}'\rho_{41}'}{Q_+(\rho')},\quad
    \rho_{42}^{+,1}=
\frac{\rho_{22}'\rho_{23}'+\rho_{33}'\rho_{32}'}{iQ_+(\rho')}.
\end{equation}
These terms do not contribute to the diagonal ones in higher-order repetitions. 
A second iteration with both two-qubit operations $M_\pm$ results again in an $X$-state of the form
\begin{equation}
\rho^{+,1}
\xrightarrow{M_\pm}\rho^{+,2}=
	\begin{pmatrix}		\rho^{+,2}_{11}&0&0&\rho^{+,2}_{14}\\	0&\rho^{+,2}_{22}&\rho^{+,2}_{23}&0\\
	0&\rho^{+,2}_{32}&\rho^{+,2}_{33}&0\\
		\rho^{+,2}_{41}&0&0&\rho^{+,12}_{44}
	\end{pmatrix}.
\end{equation}
With this strategy, the protocol suffers a bifurcation, as the states are not the same if initially $M_-$ or $M_+$ is applied. The two possibilities can be sketched in the following way
\begin{equation}
H^{\otimes 2}\rho' H^{\otimes 2}
\rightarrow
\left\{
\begin{matrix}
\xrightarrow{M_-}\rho^{-,1}\xrightarrow{M_\pm}\rho^{-,2}\dots
\rho^{-,l}\xrightarrow{M_\pm}\rho^{-,l+1}
\\
\\
\xrightarrow{M_+}\rho^{+,1}\xrightarrow{M_\pm}\rho^{+,2}\dots
\rho^{+,l}\xrightarrow{M_\pm}\rho^{+,l+1}
\end{matrix}
\right.,
\nonumber
\end{equation}
where  
$\xrightarrow{M_\pm}$ stands for one iteration process composed of the steps {\bf (I)} to {\bf (III)} as explained in section \ref{sec:oneit}.
Assuming that $\rho'$ is the initial state, the overall success probability is given in terms of the success probabilities for each event as
\begin{equation}	 
P=P_-+P_+,\quad
P_\pm=Q_\pm(\rho')
\prod_{l=1}^{\infty}p(\rho^{\pm,l}).
 \label{eq:probHad}
\end{equation}
Which states can be purified, is computed using the relations in Eqs. \eqref{eq:conditionpsim} and \eqref{eq:conditionpsip} with both possibilities $\rho^{\pm,1}$. We will not give a definite expression, as it is lengthy and not much elucidating for general states. However, one can anticipate that with this protocol a larger number of states can be purified, as the state in Eq. \eqref{eq:rhoHadmard}
 does not require a fidelity larger than one-half, a more flexible condition is necessary according to Eqs. \eqref{eq:conditionpsim} and \eqref{eq:conditionpsip}. In Sec. \ref{sec:Numerical} we will show this is the case using a numerical simulation. 
 
\section{Second Protocol with Hadamard gates (M2H2)}
\label{sec:SecondProtocolHadamard}

There is a second approach that one can follow to take advantage of both quantum operations $M_\pm$ for an initial $X$-state. In Eq. \eqref{eq:XHadMplus}, one can note that an application of the purification step with $M_+$, does not result in a diagonal state. However, the resulting density matrix $\rho^{+,1}$ can be transformed again into an 
$X$-state with the following separable unitary gate
\begin{equation}
    \label{eq:Ggate}   G=
    \frac{\mathbb{1}+i\sigma_x}{\sqrt{2}}
    \otimes
    \frac{\mathbb{1}+i\sigma_x}{\sqrt{2}}.
\end{equation}
This separable gate leaves invariant the Bell states $\ket{\Psi^-}$ and $\ket{\Phi^-}$ and interchanges the other two as $G\ket{\Psi^+}=i\ket{\Phi^+}$, and $G\ket{\Phi^+}=i\ket{\Psi^+}$. By applying  gate $G$ to $\rho^{+,1}$, we obtain
\begin{equation}
\label{eq:XHadMplusG}
\varrho^{1,0}= G\rho^{+,1}G^\dagger=
	\begin{pmatrix}		\rho^{+,1}_{11}&0&0&-i\rho^{+,1}_{13}\\	0&\rho^{+,1}_{22}&-i\rho^{+,1}_{24}&0\\
	0&i\rho^{+,1}_{42}&\rho^{+,1}_{33}&0\\
		i\rho^{+,1}_{31}&0&0&\rho^{+,1}_{44}
	\end{pmatrix},
\end{equation}
which is again in an $X$-state form. Now, taking into account that $M_+$ and $M_-$ are complementary measurement events \cite{Gonzalez2019}, we can exploit this by applying recursively the protocol with $M_+$ until the first $M_-$ event occurs. Then, the state is transformed to a Bell diagonal form and one can continue employing both cases $M_\pm$.
The procedure is sketched in the following way 
\begin{equation}
\label{eq:sketchM2H2}
\begin{matrix}
\varrho^{0,0}&
\xrightarrow{M_-}\varrho^{0,1}\xrightarrow[]{M_\pm}\varrho^{0,1}\dots
\\[3pt]
    M_+\big\downarrow H^{\otimes 2}G
    \\[3pt]
    \varrho^{1,0}
    &\xrightarrow[]{M_-}\varrho^{1,1}\xrightarrow[]{M_\pm}\varrho^{1,2}\dots
  \\[3pt]
    M_+\big\downarrow H^{\otimes 2}G
  \\[3pt]
    \varrho^{2,0}
    &\xrightarrow[]{M_-}\varrho^{2,1}\xrightarrow[]{M_\pm}\varrho^{2,2}\dots
  \\[3pt]
    M_+\big\downarrow H^{\otimes 2}G
  \\[3pt]
  \vdots & \vdots
\end{matrix}
\end{equation}
In this case, $M_+\big\downarrow G$ represents  one iteration of the process composed of the steps {\bf (I)} to {\bf (III)}  in Sec. \ref{sec:oneit} followed by an application of the separable gate $H^{\otimes 2}G$ to the obtained density matrix: First, $G$ brings the state to an $X$-state form, and then $H^{\otimes 2}$ transforms it into a block diagonal form as in
\eqref{eq:rhoHadmard}.
The initial state is  $\varrho^{0,0}=H^{\otimes 2}\rho'
H^{\otimes 2}$ as in Eq. \eqref{eq:rhoHadmard}. The overall success 
probability in this case can be computed as
\begin{equation}
\label{eq:probM2H2}
P=
\sum_{k=0}^\infty
    \left[
    q_-(\varrho^{k,0})
    \prod_{i=1}^\infty p(\varrho^{k,i})
    \right]
     \prod_{j=0}^{k-1} 
    q_+(\varrho^{j,0})
\end{equation}
with the convention that $\prod_{j=0}^{-1}X_j=1$. 
The term inside the square brackets in the previous equation, represents the probability to purify accumulated in the rows in \eqref{eq:sketchM2H2}. The term outside the brackets represents the probability to reach the $k$-th row. 
It might appear cumbersome at first glance, however, we will show that this formula can be simplified and will be useful for maximally entangled mixed states.

\section{Maximally entangled mixes states}
\label{sec:MEMS}
 
In this section, we employ the purification protocol for a particular class of states known as maximally entangled mixed states or MEMS. Ishizaka and Hiroshima studied in Ref. \cite{Ishizaka2000} how  increasing the degree of mixture of states limits the amount of entanglement that would be generated by the application of unitary transformations. For this reason, they proposed a class of mixed states in two-qubit systems. These states show the following property: for a given value of purity, $\mathcal{P}={\rm Tr}(\rho^2)$ they reach a maximum value of entanglement. 

MEMS possess the highest possible amount of entanglement for a given degree of  mixture. Up to local unitary transformations, that do not change purity and entanglement, they can be parametrized by a single variable and separated into two categories. Following  Ref. \cite{Buzek2005}, type $I$ states can be expressed as
\begin{eqnarray}
	\label{eq:mems1}
	{\varrho}_I=C\ket{\Phi^+}\bra{\Phi^+}+(1-C)\ket{01}\bra{01},
\end{eqnarray}
for $C$ $\in$ $\left[2/3,1\right]$, and defining 
$\alpha_\pm=(2\pm3 C)/6$, type $II$ states can be written as
\begin{equation}
	\label{eq:mems2}
	\varrho_{II}=
	\alpha_+
	\ket{\Phi^+}\bra{\Phi^+}
	+
	\alpha_-
	\ket{\Phi^-}\bra{\Phi^-}+
	\frac{1}{3}\ket{01}\bra{01}
\end{equation}
for $C$ $\in$ $[0,2/3]$.  The concurrence for the two types is actually $C$.
We  calculate the purity for both types of states and obtain the concurrence as a function of the purity 
\begin{equation}
\label{eq:CPMEMS}
    C=\left\{
    \begin{matrix}
 \left(1+\sqrt{2\mathcal{P}-1}\right)/2,&
        \mathcal{P}\in [5/9,1]\\
        \sqrt{2\mathcal{P}-2/3},&\mathcal{P}\in [1/3,5/9]
    \end{matrix}\right. .
\end{equation}
This expression defines the border of a region in a concurrence-purity  (CP) diagram, as
these states delimit the physically possible states, see e.g. Fig. \ref{fig:dencp}, or Fig. $1$ in \cite{Ishizaka2000}.

\subsection{Purification with DEJMPS and M2 protocols}

Let us analyze the DEJMPS protocol for this type of states. The first step is to discard the off-diagonal elements in the Bell basis, as would result from applying appropriate twirling operations \cite{Bennett1996b}.
In the case of MEMS in Eqs. \eqref{eq:mems1} and \eqref{eq:mems2}, only one projector changes in the following way
\begin{align*}
		\ketbra{01}{01}\to\frac{1}{2}\ket{\Psi^+}\bra{\Psi^+}
		+\frac{1}{2}\ket{\Psi^-}\bra{\Psi^-}.
\end{align*}
Hence, the concurrence is diminished to $2C-1$ for type $I$ states and to $C-1/3$ for type $II$, both of  which
are smaller than the initial value $C$.
In particular, for states with initial concurrence smaller than $1/3$, entanglement is completely lost. For $C>1/3$ the fidelity $\alpha_+$
with respect to $\ket{\Phi^+}$ is larger than one-half even after the twirling procedure. Therefore, any state with $C>1/3$ can be purified with DEJMPS protocol. 

 The core protocol M2 introduced in Sec. \ref{sec:MainProtocol} is also able to purify these types of states in the same fashion as DEJMPS protocol. This can be noted from the fact that MEMS are $X$-states in the Bell basis, and the map of our protocol for the diagonal elements is essentially the same as noted in Sec. \ref{sec:BellDiagonal}. The great benefit in our case is that we  are not required to bring those states into diagonal form. 

\subsection{Purification with M2H2 protocol}

As previously mentioned,
MEMS given in Eqs.  \eqref{eq:mems1} and \eqref{eq:mems2} are already $X$-states. Therefore, one can bring them to diagonal form using the M2H or M2H2 protocol introduced in Secs. \ref{sec:ProtocolHadamard} and \ref{sec:SecondProtocolHadamard}. Even though the M2 protocol is already able to purify those states, 
the performance increases using the protocols M2H or M2H2 assisted with Hadamard gates. Here we will concentrate on explaining the case of the M2H2 protocol.
For type $I$ MEMS, by applying two Hadamard gates at each node one obtains
\begin{align}
H^{\otimes 2}\varrho_I H^{\otimes 2}&=C\ketbra{\Phi^+}{\Phi^+}+(1-C)\ketbra{\varphi}{\varphi}, \\
\ket\varphi&=\frac{\ket{\Phi^-}-\ket{\Psi^-}}{\sqrt2}. \nonumber
\end{align}
These local single-qubit gates do not alter the entanglement nor the degree of mixture in the states. 
It follows from \eqref{map2} that one iteration of the protocol with $M_-$ is required to bring this type $I$ state to the Bell state $\ket{\Psi^+}$ with
success probability $C^2/2$. The fact that these  states can be purified in one step was already noted 
in \cite{Bennett1996b} using a CNOT gate and in  \cite{Torres2016} using the operation $M_-$. This is true for any value 
of the concurrence $C \in (0,1]$, i.e., not only those representing type $I$ MEMS. 

As for  type $II$ MEMS, one can show that they can also be purified with success probability $C^2/2$. This does
not happen in a single iteration as we will now explain.  Type $II$ MEMS $\varrho_{II}$ are in  an $X$-state form as $\rho'$ in Eq. \eqref{eq:XStates}. The only nonzero
elements are given by $\rho'_{11}=\rho'_{14}=\rho'_{41}=\rho'_{44}=1/6$, 
$\rho'_{22}=\alpha_-$, and $\rho'_{33}=\alpha_+$. 
From Eqs. \eqref{eq:rhodiagHad} and
\eqref{eq:probHadMs}, it follows that after the bilateral application of the Hadamard gates, the 
protocol succeeds with an application of $M_-$ with probability 
$q_-(H^{\otimes 2}\varrho_{II}H^{\otimes 2})=(\alpha_++\alpha_-)^2/2=2/9$ and the output state is the rank-two state
 \begin{equation}
\rho=
\frac{9}{2}\alpha_+\alpha_-
\ket{\Phi^-}\bra{\Phi^-}
+9\frac{\alpha_-^2+\alpha_+^2}{4}\ket{\Psi^+}\bra{\Psi^+}
 \end{equation}
whose  concurrence is now given by $9/4(\alpha_+-\alpha_-)^2=9C^2/4$. This is a rank-two state in the Bell basis, and according to Sec. \ref{rank-two}, it
 can be purified with a success probability equal to its concurrence $9C^2/4$,  and given the fact that it
 was brought to this state from $\varrho_{II}$ with probability $2/9$, the overall success probability is given by 
 $C^2/2$. Even though the process requires, in principle, an infinite amount of iterations,  it is still successful with the same probability as for the state $\varrho_I$, which requires only one iteration.  Nevertheless, it is remarkable that both types of MEMS can, at least, be purified with the same functional dependence on their concurrence and that this can be analytically demonstrated. Up to this point, we have only analyzed the M2H2 protocol when $M_-$ is obtained in the first iteration, i.e., the first row in \eqref{eq:sketchM2H2}.

Now, let us consider the cases when $M_+$ is obtained in the first and possible successive iterations, as these events contribute to improving the purification procedure. From Eqs. \eqref{eq:rhodiagHad}, \eqref{eq:rhooffdiagHad}, and \eqref{eq:XHadMplusG}, it can be noted that the $\rho^{1,0}$ is again in  the form of a MEMS. In particular, for the type $I$, the resulting state is now parametrized by a new concurrence 
$C_1=C^2/2p_0$, where $p_0=Q(\rho')=2(1-C)^2+C^2$ is the probability to reach this state. In \eqref{eq:sketchM2H2}, this corresponds to the step that takes the process to the second row. If after this step, the next iteration of the protocol succeeds with $M_-$, then one knows that it will be purified with probability $p_0C_1^2$. It follows that all states $\varrho^{l,0}$ in the first column  in \eqref{eq:sketchM2H2} will be MEMS. The concurrence can be evaluated analytically in an analog fashion compared to the procedure in Sec. \ref{sec:BellDiagonal} for rank two states. The result is given by
\begin{equation}
    C_l=\left[
    2^{2^l-1}(1/C-1)^{2^l}+1
    \right]^{-1},
\end{equation}
where $C_0=C$ is the initial concurrence, that for this case of type $I$ is in the interval $[2/3,1]$. Furthermore, by noting that
the success probability to climb down in \eqref{eq:sketchM2H2} is given by $p_l=C_l^2/2C_{l+1}$, and that in each row the 
purification is achieved with probability  
$C_l^2/2$, it is not hard to realize that the sum of all probabilities gives the 
complete success probability of purifying type $I$
MEMS that is given by
\begin{equation}
\label{eq:probMEMSI}
    P_I
    =
    \sum_{l=0}^\infty\frac{C}{2^{l+1}}
    \prod_{j=0}^lC_j.
\end{equation}
This is the expression for MEMS $I$ of the general probability in \eqref{eq:probM2H2} that was simple to calculate given the fact that one knows from the initial discussion in this section that the probability in the square brackets in \eqref{eq:probM2H2} is simply given by $C_l^2/2$.

One can proceed in a similar fashion with type $II$ MEMS, where $C\in[0,2/3]$. In this case, the MEMS form is retained after successful iterations of the protocol with $M_+$ with concurrence after each step given by
 $   \tilde C_l=\left(
   3/2
    \right)^{2^l-1}C^{2^l}.$
The success probability $Q_+(\rho')=1/3$ and it remains the same for succesive iteration with $M_+$. Therefore, the overall success probabiliy to purify type $II$ MEMS can be written as
\begin{equation}
\label{eq:probMEMSII}
    P_{II}=\frac{1}{2}\sum_{l=0}^\infty\frac{1}{3^l} 
    \left(
   \frac{3}{2}
    \right)^{2^{l+1}-2}C^{2^{l+1}}.
\end{equation}
It is remarkable that a relatively simple closed expression can be ontained for the success probability of both type of states. 

\begin{figure}
\includegraphics[width=0.47\textwidth]{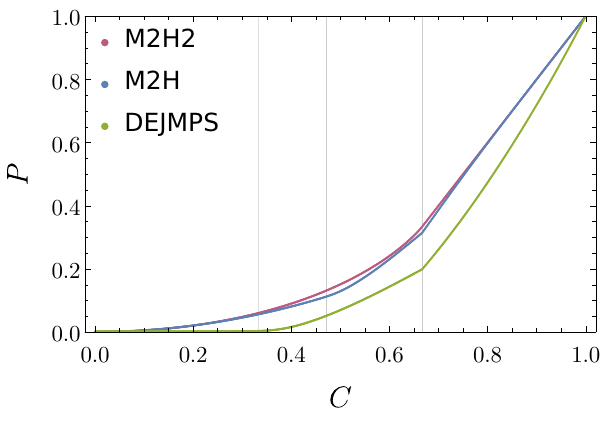}
\caption{ Success probability 
 of the purification process of MEMS type I and II as a function of the initial concurrence $\alpha$. Vertical lines are presented at $2/3$ to mark the separation between MEMS type I and II; at $\sqrt{2}/3$ below which only the operation $M_-$ is able to purify; and at $1/3$ the minimum concurrence of a purifiable MEMS with the DEJMPS protocol. }
\label{fig:num_qubits_mems}
\end{figure}

In Fig.\ref{fig:num_qubits_mems} we plot the success probability to purify all MEMS with protocol M2H2. For this,  we have employed expressions in \eqref{eq:probMEMSI} and \eqref{eq:probMEMSII}. For comparison, we have numerically evaluated the success probability using M2H protocol and DEJMPS protocol. It can be noted that the success probability with our protocols outperforms DEJMPS protocol, with slightly higher performance of the M2H2 protocol.

 As a result of this implementation of the Hadamard gate $H$, we have  found that it is possible to obtain Bell diagonal states by avoiding additional operations such as twirling \cite{Bennett1996b}. To obtain these states one needs to bring a completely arbitrary state to an $X$-state with an iteration of our protocol \cite{Torres2016}. Then, we use the Hadamard gate $H$ and finally apply one more iteration of the protocol. This procedure allows us to obtain Bell diagonal states that do not degrade the entanglement and this is less demanding in contrast to twirling operations.  These types of diagonal states are indispensable in the DEJMPS protocol.

\section{Other rank-three states}
\label{sec:Rank3States}

\label{parm}
In this section we extend our analysis to a class of rank-three states that contain all MEMS as a particular case. For this purpose we employ the incoherent superposition of a general density matrix in the two-dimensional subspace formed by the states $\ket{00}$, $\ket{11}$, and a  separable state orthogonal to them, particularly $\ket{01}$. Four parameters are needed to characterize such state that we choose to write in the following form
\begin{equation}
	\label{eq:mg}
	\rho=\frac{w+u}{2} \ket{+}\bra{+}+\frac{w-u}{2}\ket{-}\bra{-}
	+(1-w)\ket{01}\bra{01},
\end{equation}
where $|u|\le w\le 1$. The states in the former equation depend on the other two parameters and are given by 
$\ket{+}=\cos\theta\ket{00}+
e^{i\phi}\sin\theta\ket{11}$, and $\ket{-}=e^{-i\phi}\sin\theta\ket{00}-\cos\theta\ket{11}$, with $\theta\in[0,\pi/2]$, and $\phi \in [0,2\pi)$. For $\theta=\pi/2$,  
$\ket{\pm}$ are maximally 
entangled states, and additionally when $\phi=0$ they correspond to the Bell states $\ket{\Phi^\pm}$.

\begin{figure}
\includegraphics[width=0.49\textwidth]{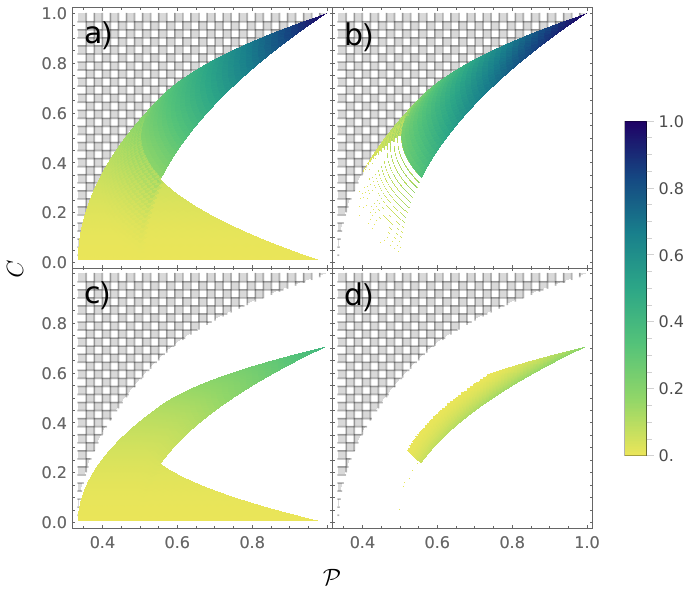}
	\caption{Phase diagram purity vs concurrence for $\theta=\pi/{2}$, $\phi=0$ and $\theta=\pi/4$, $\phi=1.3$ and number of iterations required to purify states of the form (\ref{eq:mg}).}
	\label{fig:dencp}
\end{figure}

It is not hard to realize that the concurrence and purity of this family of  states is given by 
\begin{equation}
	C=|u|\sin\theta, \quad \mathcal{P}=\frac{u^2+w^2}{2}+(1-w)^2.
\end{equation}
Note that these quantities are independent from $\phi$.
Not all combinations between $C$ and $\mathcal{P}$ are possible for these states, as the condition $|u|\le w\le 1$ has to be fulfilled. Taking that into account, one can find that
the region covered on the CP-plane for fixed $\theta$ is determined by the following inequalities
 \begin{align}
       \label{eq:CPregions}
     &0\le |u|\le \sqrt{2 \mathcal{P}-\frac{2}{3}},
     \quad \mathcal{P}\in \left[\frac{1}{3},\frac{5}{9}\right]\\ \nonumber
     &\sqrt{2 \mathcal{P}-1}\le |u|\leq \frac{1}{2} \left(\sqrt{2 \mathcal{P}-1}+1\right),\quad
     \mathcal{P}\in \left[\frac{5}{9},1\right]\\ \nonumber
     &0\le |u|\le \frac{1}{2} \left(1-\sqrt{2 \mathcal{P}-1}\right),\quad
     \mathcal{P}\in \left[\frac{5}{9},1\right],
 \end{align}
remembering that $C=|u|\sin\theta$. These regions correspond to the colored regions in  Fig. \ref{fig:dencp} a) and c)  for $\theta=\pi/2$ and $\theta=\pi/4$, respectively.  For $\theta=\pi/2$, the upper 
boundary of these inequalities coincide with the relation between purity and concurrence of the MEMS shown in Eq. \eqref{eq:CPMEMS}.

In the Bell basis, the states in Eq. \eqref{eq:mg} have the following nonzero matrix elements 
\begin{align}
&\rho_{jj}=\frac{w}{2}+(-1)^j\,\frac{u}{2}\sin\theta\cos\phi, \quad j\in\{2,3\}
\nonumber\\
&\rho_{23}=\rho_{32}^\ast=\frac{u}{2}\left(\cos\theta-i\sin\theta\sin\phi\right)
\nonumber\\
&\rho_{11}=\rho_{14}=\rho_{41}=\rho_{44}=\frac{1-w}{2}.
\end{align}
As all other elements are zero, this corroborates that these are $X$-states for any choice of the parameters. Therefore, these are good candidates to be purified with our protocols, specially the M2H2 protocol, as we have seen that analytical calculations are possible in that setting. In contrast, note that for $\phi=0$, all diagonal elements are smaller than one-half, therefore, it is not suitable for purification with the DEJMPS protocol.

After one iteration using M2H2 protocol with $M_-$, one obtains  only  the following two nonzero elements in the output state
\begin{align}
\rho_{44}=\frac{w^2-u^2\cos2\theta}{2\tilde p},\quad \rho_{22}=\frac{w^2-u^2}{2\tilde p},
\end{align}
with probability
$
\tilde p=w^2-u^2(1+\cos^2\theta)/2
$.
As this is rank two state in the Bell basis, its concurrence is easily evaluated as
$C=u^2\sin^2\theta/2\tilde p$.
From Sec. \ref{sec:BellDiagonal} we know that these type
of rank-two states can be purified with success probability $C$,
this implies that the family of rank-three states that we have considered in this case can be purified with success
probability of at leas $C^2/2$,
the same functional dependence initially found for MEMS. Note that the family of states in Eq. \eqref{eq:mg} parameterize any $X$-state, whenever $\rho_{11}=\rho_{44}=\rho_{14}$, and therefore any $X$ state of this form can be purified at least with probability given by its concurrence squared divided by two.

Let us now consider the situation when $M_+$ is applied in the first iteration (remember that this happens randomly). From Eqs. \eqref{eq:XHadMplus}, \eqref{eq:rhodiagHad}, and \eqref{eq:rhooffdiagHad} one can note that the state $\rho^{1,0}$ in 
\eqref{eq:sketchM2H2} is again an $X$-state
of the form in Eq. \eqref{eq:mg}. Therefore, it can be purified with success probability equal its concurrence squared divided by two. We will not attempt to calculate all these processes analytically, however, this already gives us a simple fashion to calculate the success probability numerically. It follows that, using Eq. \eqref{eq:probM2H2}, the term in the brackets is the concurrence of each iteration squared divided by two, and it only remains to numerically compute the product of probabilities.  

In Fig.  \ref{fig:dencp} we present the success probability of purifying  $\rho$ in Eq. \eqref{eq:mg} visualized in the CP-plane. The gridded region corresponds to unphysical combinations of concurrence and purity, and its lower border to values of the MEMS. The white region in panels a) and c) coincides with impossible combinations of $C$ and $\mathcal{P}$ for the states in \eqref{eq:mg}. The accesible region is given by the inequalities in \eqref{eq:CPregions}.
Panels
a) and c) correspond to the M2H2 protocol, whereas b) and d) to the DEJMPS protocol.
In panels a) and b), where $\theta=\pi/2$, $\phi=0$, one can note that both protocols perform similarly for large values of concurrence, however, many states are missed by DEJMPS compared to M2H2. This happens due to the strict conditions to purify with the DJEMPS procedure, where a fidelity greater than one-half with respect to a Bell state is needed. The situation is even more favorable to the M2H2 protocol when comparing panels c) and d), where $\theta=\pi/4$, $\phi=1.3$. Here one can note a diminished success probability in the DJEMPS case, but more importantly a very small portion of the states in the CP plane are purifiable using this scheme. In comparison, any state of the type considered in this section can be purified with the M2H2 protocol.
The sensitivity on $\phi$ to purify a state in DEJMPS protocol makes evident that this scheme specializes on purifying entanglement encoded in Bell states. For instance, any state with $\phi=\pi/2$ cannot be purified with this scheme, as there is no fidelity larger than one half with respect to any Bell state. In contrast, our purification method is insensitive to this parameter and can exploit the entanglement of other maximally entangled states.

\section{Numerical analysis with random states}
\label{sec:Numerical}

In this section, we extend our analysis to include random states. In order to do so, it is necessary to determine a systematic way of producing an ensemble of random states. 
We use the method presented in Refs. \cite{Zyczkowski_2001, Zyczkowski2011}, when one considers a pure state in an extended Hilbert space of dimension $4n_r$ as
\begin{equation}
\ket{\Psi}=\sum_{j=1}^4
\sum_{k=1}^{n_r} D_{j,k}\ket{j}_Q\ket{k}_A,
\end{equation}
where $\{\ket{j}_Q\}_{j=1}^4$ can be the computational basis of the two-qubit system, and with ancillary orthonormal states $\ket{k}_A$. Furthermore, the normalized probability amplitudes $D_{j,k}$ are taken as complex random variables from a Gaussian distribution with zero mean and unit variance.
\begin{figure}
\includegraphics[width=0.47\textwidth]{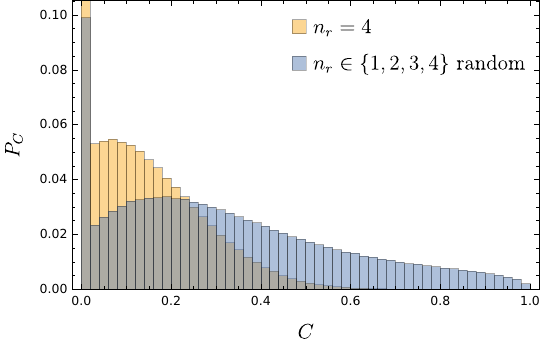}
	\caption{Probability $P_C$ of finding a state with a concurrence around a value $C$. The histograms were obtained using one million random states distributed in thirty intervals. 
    }
\label{fig:Hist}
\end{figure}
Taking partial trace of over the ancillary degrees of freedom, one can obtain the two-qubit system density matrix, whose matrix representation  is given by
\begin{equation}
\rho=DD^\dagger/{\rm Tr}(D D^\dagger).
\end{equation}
Note that the probability amplitudes can be considered as the elements  of the matrix $D$ with dimension $4\times n_r$. Only for $n_r\ge4$, one obtains a density matrix of rank four. However, in all those cases, most of the produced density matrices possess a low value of concurrence. In Fig. \ref{fig:Hist}, we present the probability of finding a density matrix with concurrence in an interval around $C$ for two different ways of selecting $n_r$ as indicated by the legend. The histograms were produced each using one million realizations of the matrix $D$ and using thirty intervals or bins for the horizontal axis. One can note that for $n_r=4$, there is a vanishing probability of finding states with concurrence larger than $0.6$. For this case, the mean value of the concurrence is also low and is given at approximately $0.126$. Before explaining the second histogram in blue color, let us comment that for $n_r<4$, the resulting density matrices increase their concurrence on average, however, their rank is given by $n_r<4$. To obtain values of the concurrence close to one with nonvanishing probability, and in addition, still consider the possibility of generating density matrices of a full rank, we choose to generate the ensemble where the integer $n_r$ is randomly varied between $1$ and $4$. The resulting probability distribution is also given in blue color in Fig. \ref{fig:Hist}, where one can note that close to unit concurrence is also possible to attain. We do not investigate the resulting probability distribution of this procedure, nor concentrate on their general properties. Here, we only take it as a systematic way of generating random density matrices that allow us to test our purification protocols in the whole range of the concurrence, i.e., $C\in (0,1)$.

\begin{figure}
\includegraphics[width=0.47\textwidth]{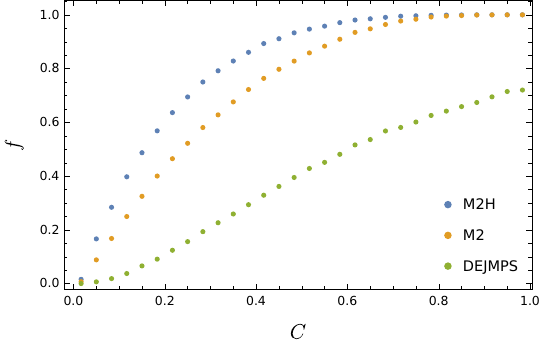}
\caption{Fraction $f$ of purifiable states from the total of one million realizations around a given value of the concurrence $C$ for three of the purification protocols as indicated by the legend.}
\label{fig:fractionnum}
\end{figure}
\begin{figure}[t!]
\includegraphics[width=0.47\textwidth]{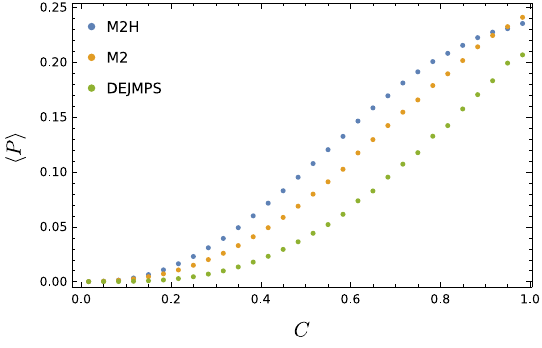}
\caption{Average success probability $P$ to be purified with each protocol, as indicated by the legend, around a certain value of the initial concurrence $C$ . Only the fraction of purifiable states over one million realizations is considered.}
\label{fig:probnum}
\end{figure}
Having explained the procedure to obtain  random states, we now proceed to test the performance of the presented purification protocols. For this purpose, we have generated an ensemble of one million random density matrices, from which we extract two quantities as a function of the initial concurrence: the fraction $f$ of purifiable states for each protocol that is presented in Fig.\ref{fig:fractionnum}, and the average success probability of each protocol presented in Fig. \ref{fig:probnum}. We test two of our protocols, namely M2 and M2H, and compare them with the DEJMPS protocol. To calculate $f$ for $M2$ protocol, we verify if conditions \eqref{eq:conditionpsim} are met for every realization. We sum all positive cases and divide them by the total number of repetitions to obtain $f$. Analogously, to calculate this number for the M2H protocol, we use again conditions \eqref{eq:conditionpsim}, but for the two possible outcomes of the second iteration of the protocol: $\rho^{\pm,1}$ given
in Eq. \eqref{eq:XHadMminus} and \eqref{eq:XHadMplus}. For the DEJMPS protocol, one has to check whether the input state has a diagonal entry larger than $1/2$ in the Bell basis.  As for the average success probability $\langle P\rangle$, we calculate the success probability of each run, sum all the results and divide by the total number of realizations. In the case of the $M2$ protocol, we use directly the expression in Eq. \eqref{eq:probM2}, whereas for the M2H protocol we use \eqref{eq:probHad} multiplied by $q_-(\rho)$, as  the results in Sec. \ref{sec:ProtocolHadamard} were calculated for initial $X$-states. As the map for DEJMPS protocol is equivalent to our map in Eq. \eqref{map2} for the diagonal entries, we use the expression in Eq. \eqref{eq:probM2X} for each generated density matrix after the elimination of all  off-diagonal elements.

From Fig. \ref{fig:fractionnum}, one can note that both of our protocols can purify a considerably larger amount of states compared to DEJMPS protocol. This is because the restriction to purify in our case is more relaxed. The protocol assisted by Hadamard gates performs better than the core M2 protocol. From Fig. \ref{fig:probnum}, we can also note an increase in the performance of both of our protocols, as the average success probability is larger in both cases compared to the result of DEJMPS protocol. This is the case even when computing this quantity over a considerably larger amount of states.

\section{Conclusions}
\label{sec:Conclusions}

In this paper, we have introduced three recurrence entanglement purification protocols and studied their performance in terms of success probabilities for different types of noisy entangled states. Analytical results are obtained  based on investigations of rank-two states in the Bell basis, where we show that these states are purifiable with success probabilities either being equal to their concurrence or the concurrence squared. The difference depends on the chosen convex combination. In particular, we have analytically shown that any maximally entangled mixed state (MEMS) can be purified with a success probability of at least its concurrence squared divided by two, using partial outcomes of one of our purification protocols. Using the complete protocol, we find a higher success probability that tends to unit success probability with the initial concurrence. Both of our protocols present a larger success probability than the DJEMPS protocol that in addition is unable to purify states with concurrence smaller than $1/3$. Based on the form of MEMS, we extend our investigation to more general rank-three states, where we also show analytically that purification can be achieved at least with a success probability of one-half the initial concurrence squared. Furthermore, we use this case to exemplify and explain how working with Bell diagonal states poses a handicap to the purification process, as it focuses on entanglement in the Bell basis, disregarding the entanglement in other maximally entangled states. The introduced protocols do not suffer from this problem and are able to purify entanglement not only encoded in the Bell basis.

Our protocols work for general input density matrices without the requirement of twirling or depolarizing operations to bring them into a diagonal form. During the purification steps, we can transform all of them into $X$-states, a form that is maintained for two of our protocols, however, we also show another one that can bring the states into Bell diagonal form using separable gates between the purification steps that do not diminish the entanglement. We test our protocols for randomly generated states and observe that a considerably larger amount of states are purifiable using our methods in contrast to the pioneering DJEMPS protocol. The fraction of purifiable states with respect to the total number of states tends to be one with our methods in contrast to DEJMPS which is less than $0.8$. We also show that the average success probability for random states is larger in our case and it tends to $0.25$ for unit concurrence in contrast to $0.22$ of the DEJMPS case. With this investigation, first, we demonstrate that efficient recurrence purification protocols do not necessarily require a CNOT gate, and in contrast, they can operate with more general operations resulting for example from two-qubit entangling measurements. In our opinion, this shows that entangling operations native to a chosen experimental platform might offer alternative or more effective solutions than abstract, sometimes hard-to-implement gate operations. Secondly, without using depolarizing  or twirling operations, which is still considered in recent developments of this topic \cite{Krastanov, Duer2021.2}, the initial loss of entanglement can be avoided and thus a larger class of quantum correlations can be exploited. Finally, we hope that the presented ideas motivate new directions of investigation of recurrence entanglement purification procedures.

\begin{acknowledgments}
This work was supported by CONAHCYT research grant CF-2023-I-1751, and from BUAP Project No. 100527172-VIEP2023.
JZB acknowledges support by AIDAS - AI, Data Analytics and Scalable Simulation, which is a 
Joint Virtual Laboratory gathering the Forschungszentrum J\"ulich (FZJ) and the French Alternative Energies and Atomic Energy Commission (CEA). 
\end{acknowledgments}

\appendix
\begin{widetext}
\section{Probability distribution of the recurrence protocol}
\label{AppendixA}

In this appendix, we consider the probabilistic description of a sequentially applied purification protocol on several input two-qubit states and determine the mean of the output states. Each iteration of the protocol is a Bernoulli trial with a success probability depending on its position in the chain. In the following we introduce step-wise the discrete probability distribution of the whole protocol with identical two-qubit states used as input. The number of the input two-qubit pairs is $n=\lfloor N/2 \rfloor$. If we have only one 
iteration with success probability $p_1$ then we count the number $k_1$ of the output two-qubit states. This is a Bernoulli trial with the binomial distribution of $k_1$ successes
\begin{equation}
 P^{(1)}(k_1) = \binom{n}{k_1}p^{k_1}_1(1-p_1)^{n-k_1}, 
\end{equation}
where $0 \leqslant k_1 \leqslant n$ and $\binom{n}{k_1}$ is a binomial coefficient. The mean is $np_1$ and the variance reads $np_1(1-p_1)$. Before the second iteration, the output two-qubit states have to paired up again and they are again identical. If the success probability of the second iteration is $p_2$ then the probability that $k_2$ two-qubit states get through in both purification rounds is  
\begin{equation}
 P^{(2)}(k_2) =  \sum^n_{k_1=2\times k_2} \binom{n}{k_1} p^{k_1}_1(1-p_1)^{n-k_1} \binom{\lfloor k_1/2 \rfloor}{k_2} p^{k_2}_2(1-p_2)^{\lfloor k_1/2 \rfloor -k_2}. 
\end{equation}
This time, the mean is
\begin{eqnarray}
 \sum^{\lfloor n/2 \rfloor}_{k_2=1} k_2 P^{(2)}(k_2) &=& \sum^{\lfloor n/2 \rfloor}_{k_2=1} k_2 \sum^n_{k_1=2\times k_2} \binom{n}{k_1} p^{k_1}_1(1-p_1)^{n-k_1} \binom{\lfloor k_1/2 \rfloor}{k_2} p^{k_2}_2(1-p_2)^{\lfloor k_1/2 \rfloor -k_2}  \\
 &=& \sum^{\lfloor n/2 \rfloor}_{k_2=1} \sum^n_{k_1=2\times k_2} \binom{n}{k_1} p^{k_1}_1(1-p_1)^{n-k_1} \underbrace{k_2 \binom{\lfloor k_1/2 \rfloor}{k_2}}_{=\lfloor k_1/2 \rfloor 
 \binom{\lfloor k_1/2 \rfloor-1}{k_2-1}} p_2 \times p^{k_2-1}_2 (1-p_2)^{(\lfloor k_1/2 \rfloor-1) -(k_2-1)} \nonumber \\ 
 &=& p_2 \sum^{\lfloor n/2 \rfloor}_{k_2=1} \sum^n_{k_1=2\times k_2} \lfloor k_1/2 \rfloor \binom{n}{k_1} p^{k_1}_1(1-p_1)^{n-k_1}  
 \binom{\lfloor k_1/2 \rfloor-1}{k_2-1}  p^{k_2-1}_2 (1-p_2)^{(\lfloor k_1/2 \rfloor-1) -(k_2-1)}, \nonumber
\end{eqnarray}
where we used $\binom{0}{0}=1$. Now, we reorganize the sums and for any fixed value of $k_1$ to obtain
\begin{equation}
 \sum^{\lfloor k_1/2 \rfloor}_{k_2=1} \binom{\lfloor k_1/2 \rfloor-1}{k_2-1}  p^{k_2-1}_2 (1-p_2)^{(\lfloor k_1/2 \rfloor-1) -(k_2-1)} =1 
\end{equation}
and thus
\begin{equation}
 \sum^{\lfloor n/2 \rfloor}_{k_2=1} k_2 P^{(2)}(k_2)= p_2 \sum^n_{k_1=2} \lfloor k_1/2 \rfloor \binom{n}{k_1} p^{k_1}_1(1-p_1)^{n-k_1}. 
\end{equation}
Now, we use the following identity 
\begin{equation}
 \lfloor k_1/2 \rfloor = \frac{k_1}{2}-\frac{1-(-1)^{k_1}}{4}, \label{eq:floor}
\end{equation}
which is valid only for natural numbers. Then, we have
\begin{eqnarray}
 p_2 \sum^n_{k_1=2} \frac{k_1}{2} \binom{n}{k_1} p^{k_1}_1(1-p_1)^{n-k_1}&=&\frac{p_2}{2} \left[ \sum^n_{k_1=1} k_1 \binom{n}{k_1} p^{k_1}_1(1-p_1)^{n-k_1}-n p_1 (1-p_1)^{n-1} \right]  \\
 &=& \frac{p_2}{2} \left[ np_1 - n p_1 (1-p_1)^{n-1} \right] = \frac{np_1p_2}{2} \left[ 1 -(1-p_1)^{n-1} \right], \nonumber
\end{eqnarray}
\begin{eqnarray}
 -p_2 \sum^n_{k_1=2} \frac{1}{4} \binom{n}{k_1} p^{k_1}_1(1-p_1)^{n-k_1}&=&-\frac{p_2}{4} \left[ \sum^n_{k_1=0} \binom{n}{k_1} p^{k_1}_1(1-p_1)^{n-k_1}- (1-p_1)^{n}-n p_1 (1-p_1)^{n-1} \right]  \\
 &=& \frac{np_1p_2}{4} (1-p_1)^{n-1} + \frac{p_2}{4} (1-p_1)^{n} - \frac{p_2}{4},  \nonumber 
\end{eqnarray}
and
\begin{eqnarray}
 p_2 \sum^n_{k_1=2} \frac{(-1)^{k_1}}{4} \binom{n}{k_1} p^{k_1}_1(1-p_1)^{n-k_1}&=&\frac{p_2}{4} \left[ \sum^n_{k_1=0} \binom{n}{k_1} (-p_1)^{k_1}(1-p_1)^{n-k_1}- (1-p_1)^{n}+n p_1 (1-p_1)^{n-1} \right] \\
 &=& \frac{np_1p_2}{4} (1-p_1)^{n-1} - \frac{p_2}{4} (1-p_1)^{n} + \frac{p_2}{4} (1-2p_1)^n. \nonumber 
\end{eqnarray}
By summing all these results together, we get
\begin{equation}
 \sum^{\lfloor n/2 \rfloor}_{k_2=1} k_2 P^{(2)}(k_2)= \frac{np_1p_2}{2} - \frac{p_2}{4} \left[1- (1-2p_1)^n \right]. 
\end{equation}
It is worth to note that
\begin{equation}
 0 \leqslant 1- (1-2p)^n \leqslant 2, \quad \forall n \in \mathbb{N} \quad \text{and} \quad p \in [0,1]. \nonumber
\end{equation}
This together with $n \gg 1/p_1$ and $p_1>0$ implies that the mean is dominated by $n p_1 p_2 /2$, or equivalently, in Eq. \eqref{eq:floor} the main contribution comes from the term $k_1/2$.

In a similar fashion, the probability that $k_m$ two-qubit states get through in $m$ purification rounds is
\begin{eqnarray}
 P^{(m)}(k_m)&=& \sum^{\lfloor n/2^{m-1} \rfloor}_{k_{m-1}=2 \times k_m} \dots \sum^{\lfloor n/2 \rfloor}_{k_2=2 \times k_3} \sum^n_{k_1=2\times k_2} \binom{n}{k_1} p^{k_1}_1(1-p_1)^{n-k_1} \binom{\lfloor k_1/2 \rfloor}{k_2} p^{k_2}_2(1-p_2)^{\lfloor k_1/2 \rfloor -k_2} \nonumber \\
 &\times& \binom{\lfloor k_2/2 \rfloor}{k_3} p^{k_3}_3(1-p_3)^{\lfloor k_2/2 \rfloor -k_3}
 \dots \binom{\lfloor k_{m-1}/2 \rfloor}{k_m} p^{k_m}_m(1-p_m)^{\lfloor k_{m-1}/2 \rfloor -k_m}.
\end{eqnarray}
Now, when $\lfloor k_j \rfloor$ ($j \in \{1,2, \dots, m\}$) is rewritten according to \eqref{eq:floor}, we systematically keep only the $k_j$ part, because the remaining terms are always less than $0.5$. This allows us to obtain
\begin{equation}
\sum^{\lfloor n/2^{m-1} \rfloor}_{k_m=1} k_m P^{(m)}(k_m)=\frac{n}{2^{m-1}}p_1p_2 \dots p_m -\delta, 
\end{equation}
where $0 \leqslant \delta \leqslant 0.5$. If $n \gg 1/(p_1 p_2 \dots p_{m-1})$ and $p_1, p_2, \dots, p_{m-1}>0$ then the mean is dominated by $n p_1 p_2 \dots p_m /2$. We recall now that  $n=\lfloor N/2 \rfloor$, which means the average two-qubit states after $m$ iterations with $N$ input qubit pairs is dominated by $\lfloor N/2 \rfloor p_1 p_2 \dots p_m /2$. Higher moments can be calculated from these discrete probability distributions, but, here, we do not investigate them. 

A main hurdle is that the output of the $i$th iteration will be the input of the $i+1$th one including a pairing among the survived states. If one considers that the two-qubit states are not identical, one has to consider in addition the pairing problem. For this,  let us assume that we have $N$ number of two-qubit states and to perform one iteration of the protocol, where one has to pair them up. If $N$ is even, then
\begin{equation}
 N_{\text{pairings}}=(N-1) \times (N-3) \times \dots \times 1 = \frac{N (N-1) (N-2) (N-3) \dots 1}{N (N-2) \dots}=\frac{N!}{(N/2)! 2^{N/2}}, \nonumber
\end{equation}
otherwise
\begin{equation}
 N_{\text{pairings}}=(N-1) \times (N-3) \times \dots \times 4 \times 3 = 3 \times 2^{\lfloor N/2 \rfloor -1} \lfloor N/2 \rfloor!, \nonumber
\end{equation}
where $\lfloor \cdot \rfloor$ is the floor function. This consideration is essential, when the two-qubit states are not identical, e.g., in pumping schemes \cite{Duer2007}.  
In the case of identical two-qubit states, this pairing problem does not play a role and thus it is not considered in the main text.

\end{widetext}

	\bibliography{manuscript}
\end{document}